\documentclass[aps,prl,twocolumn,showpacs]{revtex4}
\usepackage{graphicx}
\usepackage{amsmath,amsfonts}

\newcommand{\er}{\mathbf{r}}
\newcommand{\ee}{{\rm e}}

\begin{document}

\title{Rotonlike instability and pattern formation in spinor Bose-Einstein condensates}

\author{Micha\l{} Matuszewski}
\affiliation{Instytut Fizyki PAN, Aleja Lotnik\'ow 32/46, 02-668 Warsaw, Poland, \\
and Nonlinear Physics Center, Research School of Physics and
Engineering, Australian National University, Canberra ACT 0200,
Australia}

\begin{abstract}
We show that metastable phases of an antiferromagnetic spin-1 condensate 
in a simple model with pure contact interactions can exhibit a rotonlike minimum in the excitation spectrum.
The introduction of magnetic field gives rise to the instability of roton modes, 
which can lead to spontaneous emergence of regular periodic, polygonal, polyhedral or crystalline patterns,
as shown in numerical simulations within the truncated Wigner approximation.
An explanation of the occurrence of rotonlike instability is given based on the energy and spin conservation laws.
\end{abstract}
\pacs{03.75.Kk, 03.75.Mn, 67.85.De, 67.85.Fg}

\maketitle

Bose-Einstein condensates with spin degrees of freedom~\cite{Ho} attracted in recent years 
great interest due to the unique possibility of exploring fundamental concepts of quantum mechanics
in a remarkably controllable and tunable environment.
The ability to generate spin squeezing and entanglement~\cite{Entanglement}
makes spinor Bose gases promising candidates for 
applications as quantum simulators~\cite{QS}, in quantum information~\cite{QI}, and for precise measurements~\cite{Measurement}.
Moreover, spinor condensates were successfully used to recreate many of the phenomena of condensed matter physics
in experiments displaying an unprecedented level of control over the quantum system.
In particular,
spin domains~\cite{Domains}, spin mixing~\cite{Mixing}, and spin vortices~\cite{Ketterle_Coreless}
were predicted and observed.

The fundamental concept of a roton excitation, first introduced by Landau in the context of superfluid helium, 
is crucial for understanding of its physical properties~\cite{Feynman}. 
It is characterized by a minimum in the spectrum of excitations occurring at a finite
wavelength, $E(k) \approx \Delta + \hbar^2(k-k_0)^2/2\mu$. If the roton gap $\Delta$ can be decreased by changing the system
parameters, the softening of the roton mode can eventually lead to an instability.
This instability scenario is encountered also in many other branches of quantum physics,
including strongly correlated Fermions~\cite{fermions}, quantum Hall systems~\cite{QH}, 
and Bose-Einstein condensates with long range interactions~\cite{Softening,Dipolar_Rotons,Rydberg}.
The roton instability is characterized by unstable modes with wavevector lengths close to the roton minimum $k_0$.
It was suggested that it can lead to the emergence of the peculiar supersolid state~\cite{supersolid}, and
several other physical phenomena~\cite{Dipolar_Collapse}.

Here, we show that rotonlike instability can occur in spinor Bose-Einstein condensates
in a simple model with pure contact interactions. 
It can take place in appropriately prepared metastable states
of an antiferromagnetic spin-1 condensate \cite{Matuszewski_PS} under the influence of magnetic field.
Moreover, we show that, depending on the geometry and the trapping potential, it can lead to 
spontaneous emergence of variety of transient ordered patterns, including polygonal, polyhedral and crystalline structures. 
We show that these results can be verified in experiments with $^{23}$Na condensate, which is characterized
by very weak dipolar interactions~\cite{Na_dipolar}. We provide an explanation for the occurrence of rotonlike instability
based on energy and spin conservation laws, and demonstrate how the pattern characteristic length is determined by
the transfer of the spin energy to the kinetic energy.

We consider a dilute spin-1 BEC in a homogeneous magnetic field pointing along the $z$ axis.
We start with the Hamiltonian $\hat{H} = \hat{H}_0 + \hat{H}_{\rm A}$, where the symmetric (spin independent) part is
\begin{equation} \label{En}
\hat{H}_0 = \sum_{j=-,0,+} \int d\er \, \hat{\psi}_j^\dagger \left(-\frac{\hbar^2}{2m}\nabla^{2} + \frac{c_0}{2} \hat{n} 
+ V({\bf r})\right) \hat{\psi}_j,
\end{equation}
where the subscripts $j=-,0,+$ denote sublevels with magnetic quantum numbers along the $z$ axis $m_f=-1,0,+1$,
$m$ is the atomic mass, $\hat{n}=\sum \hat{n}_j = \sum \hat{\psi}_j^\dagger \hat{\psi}_j$ is the total atom density and 
$V({\bf r})=\frac{1}{2}m\omega_{\perp}^2(x^2+y^2)+\frac{1}{2}m\omega_{z}^2z^2$
is the external potential. The asymmetric part can be written as
\begin{equation} \label{EA}
\hat{H}_{\rm A} = \int d\er \, \left(\sum_j E_j \hat{n}_j + \frac{c_2}{2} :\hat{{\bf F}}^2:\right)\,,
\end{equation}
where $E_j$ are the Zeeman energy levels 
and the spin density is 
$\hat{{\bf F}}=(\hat{\psi}^{\dagger}F_x\hat{\psi},\hat{\psi}^{\dagger}F_y\hat{\psi},\hat{\psi}^{\dagger}F_z\hat{\psi})$
where $F_{x,y,z}$ are the spin-1 matrices and $\psi =(\psi_+,\psi_0,\psi_-)$.
The spin-independent and spin-dependent interaction coefficients are given by $c_0=4
\pi \hbar^2(2 a_2 + a_0)/3m$ and $c_2=4 \pi \hbar^2(a_2 -
a_0)/3m$, where $a_S$ is the s-wave scattering length for colliding atoms
with total spin $S$.

The linear part of the Zeeman effect induces a homogeneous rotation of the spin vector around the direction of the magnetic field.
Since the Hamiltonian is invariant with respect to rotations around the $z$ axis, we can remove this trivial effect by
introducing a frame rotating with the Larmor frequency.
We thus consider only the effects of the quadratic Zeeman shift~\cite{Matuszewski_PS}.
For sufficiently weak magnetic field we can approximate it by a positive energy shift of the $m_f=\pm 1$ sublevels $\delta
E=(E_+ + E_- - 2E_0)/2  \approx \alpha^2 E_{\rm HFS}/16$, where $E_{\rm HFS}$ is the hyperfine energy splitting at zero
magnetic field, $\alpha = (g_I + g_J) \mu_B B/E_{\rm HFS}$, 
$\mu_B$ is the Bohr magneton, $g_I$ and $g_J$ are the gyromagnetic
ratios of electron and nucleus, and $B$ is the magnetic field strength \cite{Matuszewski_PS}.

In the mean field approximation, the Hamiltonian gives rise to the Gross-Pitaevskii equations 
\begin{align}\label{GP}
i \hbar\frac{\partial \psi_{\pm}}{\partial t}&=\left[ \mathcal{L} +
c_2 (n_{\pm} + n_0 - n_{\mp})\right] \psi_{\pm} +
c_2 \psi_0^2 \psi_{\mp}^* \,, \\\nonumber
i \hbar\frac{\partial \psi_{0}}{\partial t}&=\left[ \mathcal{L} -
\delta E + c_2 (n_{+} + n_-)\right] \psi_{0} + 2 c_2
\psi_+ \psi_- \psi_{0}^* \,,
\end{align}
where $\mathcal{L}=-\hbar^2\nabla^2/2m+c_0n + V({\bf r})$.
The total number of atoms $N=\int n \,d \er$ and the total magnetization 
$M= \int F_z d \er = \int \left(n_+ - n_-\right) d \er$
are constants of motion. 

\begin{figure}
\includegraphics[width=8.5cm]{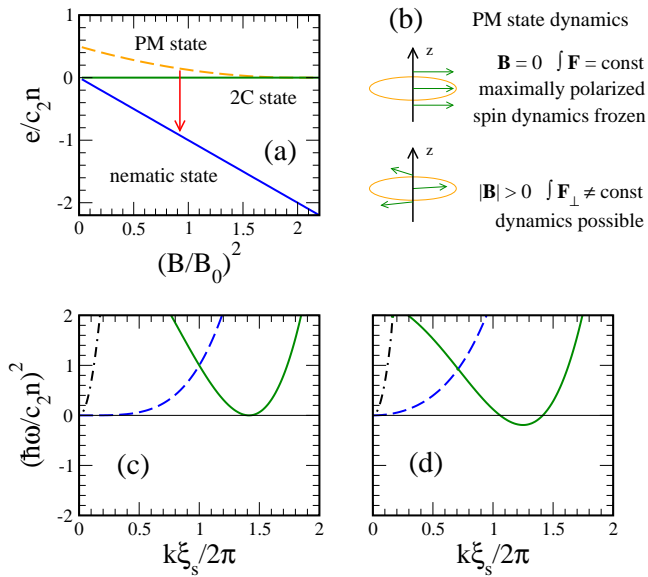}
\caption{(a) Energy spectrum of stationary states of a uniform spin-1 antiferromagnetic condensate with magnetization 
$M=0$ as a function
of the magnetic field strength. The scaling field $B_0$ depends on the density.
(b) The phase-matched (PM) phase is an excited state, however in the case $B=0$ 
it corresponds to maximal transverse polarization and its stability is ensured by spin conservation. For $B > 0$
the transverse spin is no longer conserved and instability occurs. 
(c) Excitation spectrum of the PM phase in the case $B=0$. 
The dash-dotted, dashed, and solid lines correspond to density wave, spin wave and spin quadrupole Bogoliubov modes. 
(d) The same for $B=1.16 B_0$. The minimum in the quadrupole mode gives rise to an instability characterized 
by imaginary frequencies.
The spin healing length is $\xi_s=2\pi\hbar / \sqrt{2m c_2 n}$.
}
\label{spectra}
\end{figure}

First, we describe possible phases of a condensate in a homogeneous magnetic field in the case of a vanishing potential,
$V({\bf r}) = 0$ \cite{Matuszewski_PS}.
Homogeneous stationary solutions have the form
$\psi_{j}(\er,t) = \sqrt{n_{j}} \ee^{-i\mu_{j}t + i \theta_j}$,
where $\theta_j$ are the phase shifts and $\mu_+ + \mu_- = 2\mu_0$.
These solutions are stationary in the sense that the number of atoms in each magnetic sublevel $n_j$ is
constant in time, but the relative phases may change as a result of an additional spin precession around $z$.
In Fig.~\ref{spectra}(a) we show the energy per atom in various phases of an antiferromagnetic condensate ($c_2>0$)
as a function of the magnetic field strength for $M=0$. 
Because the symmetric part of the Hamiltonian~(\ref{En}) is a constant, the relevant part of the energy $e(r)$
is defined through $H_{\rm A} =  \int d\er \, n\, e(\er)$.
Note that due to scaling properties of the Hamiltonian,
the use of renormalized variables $e/(c_2 n)$ and $B/B_0$, where $B_0$ depends on the density through the condition $\delta E|_{B=B_0} = c_2 n$,
allows to show the possible phases of spin-1 condensate in one universal graph (there is no fixed parameter except $M$).
The nematic ($\rho_0$) state is described by $\psi=(0,\sqrt{n},0)$, the two-component (2C) state
by $\psi=(\sqrt{n/2},0,\sqrt{n/2})$, while in the phase-matched (PM) state all three magnetic components are populated with the
relative phase between them equal to  $\theta=\theta_++\theta_--2\theta_0=0$
and relative populations dependent on the magnetization. In the limit of zero magnetic field, nematic and 2C states
become polar states, while PM states become ferromagnetic states~\cite{Ho}. For more details about the possible phases,
see~\cite{Matuszewski_PS}.

The PM phase is the highest excited state at low magnetic fields, 
since it is equivalent to the fully polarized ferromagnetic state at $B=0$, and this polarization 
gives a dominant positive contribution to the Hamiltonian~(\ref{EA}). Despite being an excited state, the PM phase
is stable at $B=0$ \cite{Matuszewski_AFDI} due to spin conservation, 
which ensures that the condensate remains fully polarized. However, after introducing magnetic field, 
the perpendicular part of the spin is no longer conserved
and the phase becomes unstable, as depicted in Fig.~\ref{spectra}(b).
We note that the growth rate of unstable modes of the PM phase at weak fields is proportional 
to the fourth power of the magnetic field strength \cite{Matuszewski_AFDI}, and 
the time of the development of instability may be much longer than the condensate 
lifetime.

We investigate the destabilization of the PM phase in detail by calculating the growth rate of
linear Bogoliubov modes \cite{Ho,Ueda_spin1}, 
$\psi_j=\psi_j+ \delta \psi_j$ with $\delta \psi_j=\left(u_j(t) \ee^{i{\bf k \cdot x}} + v_j^*(t)
\ee^{-i{\bf k \cdot x}} \right) \ee^{-i \mu_j t + i \theta_j}$,
where $u(t),v^*(t)\sim \ee^{i\omega t}$, $\omega$ being the
eigenfrequency of the excitation. 
In Fig.~\ref{spectra}(c,d) we present spectra of excitations
for the PM phase in absence and presence of magnetic field.
The solid line, corresponding to spin quadrupole modes \cite{Ho, Ueda_spin1}, exhibits a minimum, which in nonzero
magnetic field, Fig.~\ref{spectra}(d), gives rise to unstable modes (imaginary frequencies) localized around a finite wavelength value, 
of the order of the spin healing length $\xi_s$. Roton modes are characterized by an energy minimum at nonzero momentum
\cite{Feynman}. Here the minimum corresponds to massless, gapless ($\Delta, \mu=0$) rotons
at $B=0$. This picture is analogous to that at the point of instability in dipolar condensates \cite{Softening}. However, 
the phonon part of the spectrum at low momenta and the maxon (maximum) feature \cite{Dipolar_Rotons} are absent in the current case.

\begin{figure}
\includegraphics[width=8.5cm]{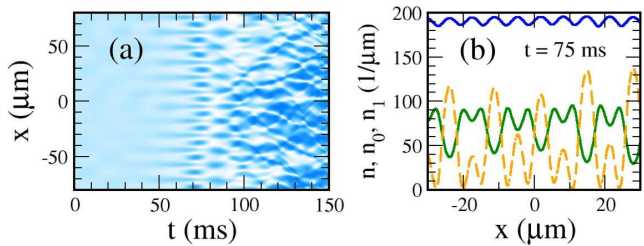}
\caption{Onset of the instability and spontaneous periodic pattern formation in a $^{23}$Na condensate in 
a quasi one dimensional regime. (a) Density of atoms in the $m_f=0$ sublevel, $n_0(x,t)$, in a ring geometry
with periodic boundary conditions at $x=\pm80 \mu$m. 
(b) Snapshot of the densities in the sublevels at $t=75$ms.
The upper solid line is the total density $n$, and the lower solid and dashed lines correspond to $n_1$ and $n_0$, respectively.
The density $n_{-1}$ is very close to $n_1$.
Other parameters are $N=3\times 10^4$,  $\omega_{\perp}=2\pi\times 10^3\,$Hz, $M=0$, $B=0.4$G, $N_0=0.17\,$N.
The excitation spectrum of the initial state corresponds to Fig.~\ref{spectra}(d).}
\label{1d}
\end{figure}

To illustrate the onset of instability in the PM phase, in Fig.~\ref{1d} we present results of numerical simulations 
of $^{23}$Na atoms confined in one-dimensional geometry with periodic (ring) boundary conditions~\cite{Matuszewski_AFDI}. 
The calculations were carried out within the truncated Wigner approximation with the initial noise filtered to remove
high spatial frequencies above $k_{\rm max}=0.2 \mu m^{-1}$. Such a choice
of initial noise gives a better agreement between theory and experiment \cite{Ueda_Defect}. 
It is clearly visible that the instability leads to spontaneous appearance of transient periodic patterns before
they dissolve into random structures. The spatial period of the  
pattern corresponds to the wavelength of the unstable quadrupole modes. We note that similar patterns were observed recently
in a $F=2$ condensate prepared in an analogous, transversely polarized state
subject to magnetic field \cite{Sengstock_Patterns}.

\begin{figure}
\includegraphics[width=8.5cm]{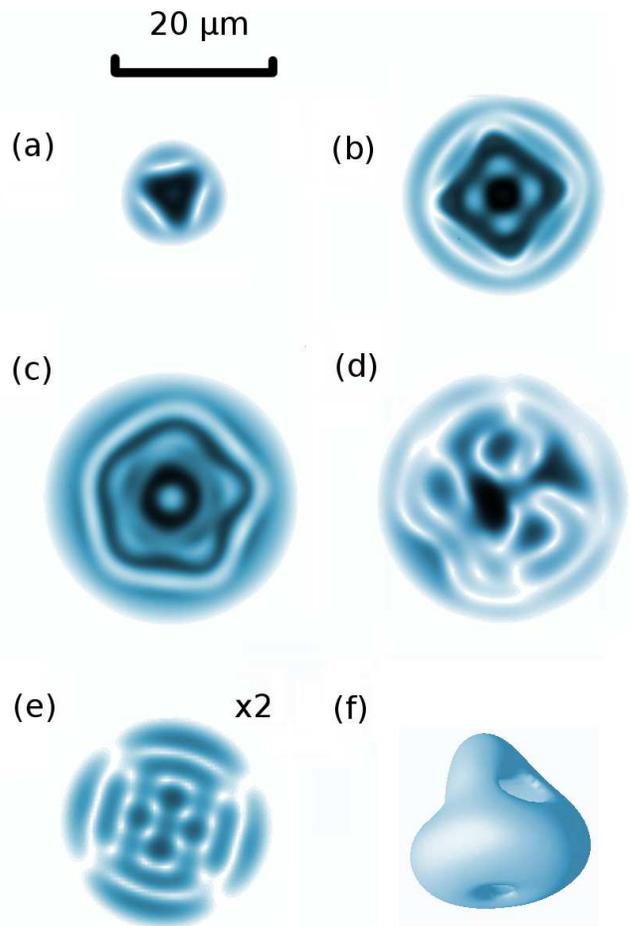}
\caption{Transient patterns formed spontaneously in a quasi two dimensional and three dimensional settings in harmonic traps.
Pictures (a)-(e) show snapshots of the density and (f) shows the isosurface of the density of atoms 
in the $m_f=0$ sublevel after a specified time of evolution.
The initial states are close to the PM state (see text), except (d) and (e), where the initial state is the 
2C state with wide instability spectrum in (d) and narrow, bounded instability spectrum in (e).
Parameters are 
(a) $N=3\times 10^4$,  $\omega_{\perp}=2\pi\times 80\,$Hz, $t=75\,$ms, 
(b) $N=16\times 10^4$,  $\omega_{\perp}=2\pi\times 50\,$Hz, $t=57\,$ms, 
(c) $N=2\times 10^5$,  $\omega_{\perp}=2\pi\times 35\,$Hz, $t=62\,$ms, 
(d) as in (c) with $t=48\,$ms, 
(e) $N=4\times 10^4$,  $\omega_{\perp}=2\pi\times 100\,$Hz, $t=46\,$ms, 
(f) $N=3\times 10^5$,  $\omega_{\perp}=\omega_{z}=2\pi\times 100\,$Hz, $t=152\,$ms.
Other parameters are $B=0.4$G except (e) where $B=0.8$G, and $\omega_{z}=2\pi\times 10^3\,$Hz in (a)-(e). 
Densities in trap centers correspond approximately to $B_0=0.4\,$G. The frame (e)
is magnified two times.
}
\label{gallery}
\end{figure}

In Fig.~\ref{gallery} we present results in two- and three-dimensional geometries with harmonic trapping potentials.
Following a typical experimental scenario, the initial state is prepared from 
the ground state 
with all the atoms in the 
$m_f=1$ component, by coherently transferring part
of the atoms to the other components and keeping the mutual phase equal to $\theta=0$. Because the density of atoms varies in
space and so the atom distribution in the PM phase, it is not possible to prepare a perfect initial state with this technique.
We prepare an imperfect PM state by setting the atom distribution to $N_+$:$N_0$:$N_-$=2:1:2,
which corresponds approximately to the average density $\langle n \rangle$.
Figures~3(a)-(c) show snapshots of the atomic density in the $m_f=0$ sublevel in the two-dimensional setting with increasing
condensate dimensions. 
The rotonlike instability gives rise to the emergence of regular patterns in similar conditions as in the untrapped case,
but only if the condensate size is larger than the pattern characteristic length, which roughly corresponds to 
the healing length for the average density $\langle n \rangle$.
The symmetry of the patterns reflects the symmetry of unstable
modes, which can change substantially depending on the ratio between the characteristic length and the condensate size.
For comparison, in Fig.~\ref{gallery}(d) we present 
irregular pattern generated from the 2C state, which has a non-roton instability spectrum,
with no lower bound for wavevectors $k$. Figure~\ref{gallery}(e) shows a crystalline pattern created
from the 2C state at a higher magnetic field, after the bifurcation with the PM state, when the spectrum becomes bounded (see below).
In Fig.~\ref{gallery}(f) we present an example of a tetrahedron-like pattern generated in a three-dimensional setting.
We note that almost perfect crystalline patterns could be generated form the PM state in the absence of trapping potential.

\begin{figure}
\includegraphics[width=8.5cm]{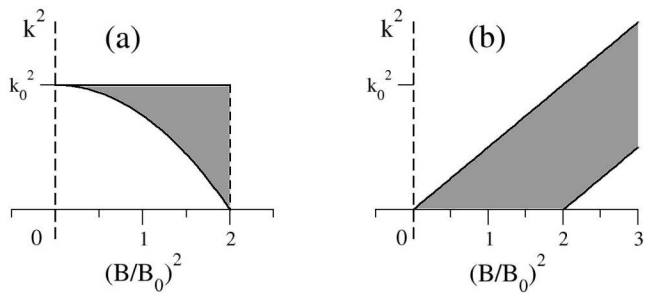}
\caption{(a) Emergence of the narrow instability spectrum (shaded) in the PM phase. 
At the instability threshold, $B=0$, unstable wavevectors are localized
around a nonzero value $k_0$.
(b) For comparison, the instability of the 2C phase has initially no lower bound for the 
wavevector length, and produces irregular patterns, see Fig.~\ref{gallery}(d). At
stronger magnetic fields, the instability spectrum becomes bounded
and gives rise to regular structures as in Fig.~\ref{gallery}(e).
}
\label{mi}
\end{figure}

The above results can be understood by considering the spectra of unstable modes in Fig.~\ref{mi}.
Both the intuitive picture and the presented numerical simulations 
suggest that the regular structures emerge when the Bogoliubov spectrum contains only a limited
range of unstable modes, with wavevector lengths 
close to the characteristic length of the generated pattern. 
This kind of spectrum appears naturally when the instability 
is a result of crossing of a roton minimum into the unstable domain.
We argue that this situation occurs in excited states whose
relaxation is prevented by a conservation law.
If the constraint can be relaxed gradually,
the instability is likely to emerge only in a narrow range of finite wavevector values.
This intuitive reasoning is supported by the following calculation of the wavelength of the most unstable Bogoliubov mode
of the PM state.
At the instability threshold, $B=0$, we choose one of the possible forms of the mean-field PM phase
$\psi_0=\sqrt{n}(\frac{1}{2},\frac{1}{\sqrt{2}},\frac{1}{2})$, and the quadrupole mode
$\delta \psi = \varepsilon (\frac{1}{2},-\frac{1}{\sqrt{2}},\frac{1}{2}) \ee^{ikx}$ without loss of generality.
To estimate the value of unstable wavevectors,
we calculate the normalized spin energy per atom 
$e_{\rm s}=\frac{1}{2}c_2 n-8\varepsilon^2c_2n + O(\varepsilon^4)$
and the kinetic energy per atom $e_{\rm k}=2\varepsilon^2\hbar^2k^2 /m + O(\varepsilon^4)$, 
assuming that a single most unstable linear mode is dominant.
Since the spin energy is converted into kinetic energy,
the energy conservation condition $\hbar^2k_0^2=4 m c_2 n$ determines the value of $k_0$, which
agrees perfectly with that of unstable modes appearing
at $B=0$ in Fig.~\ref{mi}(a).
In comparison, the instability of the 2C state, which
is initially not separated by an energy gap from the ground state, cf.~Fig.~\ref{spectra}(a),
has no lower bound for the wavevector value, see Fig.~\ref{mi}(b).
Hence the narrow bounded spectrum in the PM phase is due to finite energy gap and  
spin conservation, which inhibits relaxation of the excess energy at low magnetic fields.
This argument also allows to understand the absence of rotonlike instability in ferromagnetic condensates.
In this case, the excited nematic and 2C states~\cite{Matuszewski_PS} are characterized by zero transverse
spin, and the spin conservation does not prevent energy relaxation.

In conclusion, we demonstrated that the excitation spectrum of an antiferromagnetic condensate  
in a simple model with pure contact interactions can exhibit a rotonlike minimum.
Under the influence of magnetic field, this minimum gives rise to an instability and can lead to spontaneous emergence of 
regular periodic, polygonal, polyhedral or crystalline patterns.
Theoretical considerations suggest that the appearance of rotonlike instability is related
to spin conservation, which inhibits the relaxation of excess energy at low magnetic fields.

This work was supported by the EU project NAMEQUAM and the ARC Centre of Excellence ACQAO.

\clearpage

\end{document}